\documentclass[letterpaper, 10 pt, journal, twoside]{ieeetran} 
\usepackage{cite}
\usepackage{textcomp}
\usepackage{cleveref}
\usepackage{color,soul}
\usepackage{gensymb}
\usepackage{paralist}
\usepackage{amsthm}
\usepackage[pdftex]{graphicx}
\usepackage[cmex10]{amsmath}
\usepackage{amsfonts}
\usepackage{amssymb}

\usepackage{url}
\usepackage{color,soul}
\usepackage{setspace}
\usepackage{hyphenat}

\newcommand{\RN}[1]{%
  \textup{\uppercase\expandafter{\romannumeral#1}}%
}

\newtheorem{definition}{\textbf{Assumption}}
\newtheorem{Def}{Definition}
\newtheorem{remark}{Remark}

\IEEEoverridecommandlockouts                              

\title{A Local Polynomial Approach to Nonparametric Estimation of the Best Linear Approximation of Lithium-Ion Battery from  Multiple Datasets \\ \linespread{2}
 \footnotesize{\textcolor{blue}{ArXiV Preprint: The original article is published in  IEEE Control Systems Letters  Vol. 1, Issue: 1, pp. 182\hyp{} 187, July 2017, \\ DOI: 10.1109/LCSYS.2017.2712364}}}

\author{Rishi Relan$^{1}$, Koen Tiels$^{1}$, Jean-Marc Timmermans$^{2}$, and Johan Schoukens$^{1}$
\thanks{The authors$^{1}$ are  with  the  Dept. ELEC and the author$^{2}$ was with Dept. ETEC, MOBI group, Vrije Universiteit Brussel, Pleinlaan 2, 1050-Brussels. The author$^{2}$ is now with Agoria, Agoria AA Tower - Technologiepark 19, B-9052 Gent (Zwijnaarde), Belgium. The corresponding author can be contacted at {\tt\small rishi.relan@vub.ac.be}. 
}
}

\begin{document}
\bstctlcite{IEEEexample:BSTcontrol}

\maketitle
\thispagestyle{empty}
\pagestyle{empty}

\begin{abstract}
Battery short-term electrical impedance behaviour varies between linear, linear time-varying or nonlinear at different operating conditions. Data based electrical impedance modelling techniques often model the battery as a linear time-invariant system at all operating conditions. In addition, these techniques require extensive and time consuming experimentation. Often due to sensor failures during experiments, constraints in data acquisition hardware,  varying operating conditions and the slow dynamics of the battery, it is not always possible to acquire data in a single experiment. Hence multiple experiments must be performed.  In this  paper, a local polynomial approach is proposed to estimate nonparametrically the best linear approximation of the electrical impedance affected by varying levels of nonlinear distortion, from a series of input current and output voltage data sub-records of arbitrary length.

\end{abstract}
\section{INTRODUCTION}

\IEEEPARstart{E}{mpirical} and semi-empirical models are good alternatives to the highly complex electro-chemical, electro-thermal or thermo-chemical models to describe the short-term electrical response of the battery. If an adequate amount of training data are available which is acquired under different operating conditions, then the data-driven methods are significantly more efficient than the model-based methods in terms of computation, execution time, and memory requirements. Therefore, in a large number of investigations and applications, a good estimation of the battery's electrical impedance is obtained from the measured input and output data. 

Electrical  impedance  measurements  provide  useful information about the characteristics of a Li-ion battery \cite{thele2008development, andre2011characterization}. Electrochemical impedance spectroscopy (EIS) is the classical tool to do these measurements. It consists in the application of an electrical stimulus to the working electrode and then monitoring of its corresponding response. These experiments involve a stepwise change of frequency in the applied sinusoidal  current,  measuring  the  corresponding  sinusoidal voltage and then calculating at each frequency the electrochemical impedance. Although robust, it is an expensive, complex, and very time-consuming method. The authors in \cite{karden2000method,thele2005hybrid} performed galvanic EIS and measured impedance by adding DC current with different levels to the AC perturbation. One of the drawbacks of testing with high current levels is that, it distorts the impedance due to significant nonlinear distortions (NL). To avoid significant state of charge (SoC) changes during the test, the frequency range is also limited to a rather small band during these experiments. 

The effect of rest duration before measurement, on the estimation of impedance was investigated by \cite{andre2011characterization}, while they did not consider transient effects on the measurement during the experiment. Furthermore, the operating conditions have a significant impact on the performance and the capacity of the batteries. Experimental results show an important interaction between the electrical and thermal phenomena \cite{RishiICC2016}. The relationship between the input current and output voltage is a nonlinear function of temperature \cite{haran2002capacity} e.g. the  amount  of  energy  stored  inside  depends  largely  on  the temperature. Hence, the estimation of electrical impedance at different temperatures will result in multiple datasets.

Apart from the change in physical parameters like e.g. temperature, humidity, pressure etc, in practice there are many situations during an experiment which can effect the short term electrical dynamics of the battery that can lead to a series of sub-records of data of equal \cite{slra-consistency} or unequal lengths \cite{johan2012conc}. For example, in a long experiment, some parts in the data can have extremely poor quality due to a sensor failure or due to very large disturbances coming from other processes. Eliminating these bad parts results in a series of small sub-records of the data. Similarly, it might be impossible to measure for a very long time without interruption; e.g. due to inadequate technical capabilities of data acquisition equipment and the lack of on-board memory for storing the data. Finally, battery dynamics vary slowly therefore, a series of sub-records under similar conditions are acquired.  

Furthermore, in order to obtain a good initialization of the nonlinear model proposed in \cite{RRelan2016, RishiTCST2016} which is valid under different operating conditions, an estimation of a common nonparametric best linear approximation (BLA) from the data acquired from multiple operating conditions is needed. If, for one of these reasons, a set of shorter sub-records is available, then it is important to develop a methodology which can handle the data from multiple experiments.

\subsection{Contribution and organization of the paper}

We propose a data-driven local polynomial method (LPM) based methodology, to develop the BLA of the battery's electrical impedance from multiple input-output datasets. These datasets are either acquired at the same or at varying operating conditions e.g. different SoC levels, temperatures etc. with varying level of noise and NL. The advantage of this method over the conventional single-sine excitation methods is the reduction in the measurement time, explicit handling of NL and better handling of the leakage errors \cite{JohanIEEECSM2016}.

This paper is organized as follows: Section \ref{MultisineDef} describes the problem statement very briefly and introduces multisine excitation signals. Section \ref{Sec:BLA} describes the concept of the BLA. Nonparametric identification procedure using LPM approach is described in Section \ref{NonParam}. The procedure to obtain the BLA for all datasets is explained in Section \ref{CommBLA}. Section \ref{MeasExp} describes the experimental set-up and the measurement methodology, which is used for the acquisition of the signals. Results of experiments are presented in Section \ref{res}, and finally, the conclusions are given in Section \ref{conc}. 

\section{Broadband Excitation for Data Acquisition}
\label{MultisineDef}
The short-term voltage response of the battery to the input current load profile at a particular setting of SoC and temperature can be approximately described by the following nonlinear relationship, where $f$ is a nonlinear function which maps SoC, current $I$ and temperature $T$ to the terminal voltage $V$ at a particular instant in time. 
\begin{equation}
 V(t) \approx f(SoC(t), I(t), T(t))
\end{equation} 
Broadband signals such as multisine signals offer various advantages over random Gaussian noise signals in extracting information from dynamical systems \cite{Bayard:1993, rivera2009constrained, al2013broadband}, but during the design process, the amplitude spectrum  of  the  multisine excitation should be designed such that the equivalence between the random phase  multisine  and  the  Gaussian  random  noise  with respect to (w.r.t.) the nonlinear behaviour is always guaranteed \cite{JohanIEEECSM2016}. Hence, the  equivalence  class $E_{S_u}$ is defined, which contains  all  signals  that  are  (asymptotically)  Gaussian  distributed, and have asymptotically, for  $N \rightarrow \infty$, where $N$ is the number of excited harmonics in a multisine, the same power on each finite frequency interval. This is precisely stated in the definition below \cite{RikJohanBook2012}. 

\begin{Def} \underline{Riemannian Equivalence Signal Class $E_{S_u}$:} Consider a piecewise continuous signal $u$ with a power spectrum $S_U(j\omega)$, with a finite number of discontinuities. A random signal belongs to the Riemann equivalence class of $u$, if it obeys by any of the following statements:
	\label{def1}
\begin{inparaenum}
 \item  It is a Gaussian noise excitation signal with power spectrum $S_U(j\omega)$.
 \item It is a random multisine or random phase multisine excitation \cite{RikJohanBook2012} such that:
 \begin{equation}
 \frac{1}{N}\sum\limits_{k = k_1}^{k_2} \mathbb{E} \{|U(j\omega_{k})|^2\}= \frac{1}{2\pi}\int\limits_{\omega_{k_1}}^{\omega_{k_2}}S_U(\nu)d\nu + \mathcal{O}(N^{-1})
 \end{equation}
\end{inparaenum}
\end{Def}  where $\omega_{k} = k \tfrac{2\pi f_{s}}{N}, k \in \mathbb{N}, 0 <\omega_{k_{1}}<\omega_{k_{2}}< \pi f_s $ and $f_s$ is the sample frequency. The  frequency domain representation of the multisine signal is given by:
\begin{equation}
U_{ms}(j\omega) = \frac{1}{\pi\sqrt{N_e}}\sum\nolimits_{k_e\in\pm\mathbb{K}_{exc}}\hspace{-0.7em}{A}(k_e) \delta(\omega - \omega_{k_e}) e^{j \varphi_k} 
 \end{equation} where $\delta(\bullet)$ is the Dirac delta function, $\mathbb{K}_{exc}\subset([1,Tf_s/2]\cap \mathbb{N})$ is the discrete set of excited frequency bins, $T$ represents both the period of the multisine and the length of the measured time record, $N_e$ the number of excited frequencies, $\omega_{k_e}$ is the excited frequency and $\varphi_{k}$ $\sim \mathcal{U}[0,2 \pi[$ are the phases. Depending on the application, the amplitudes ${A}(k) \geq 0$ can be chosen arbitrarily. In addition, the signal at some of the unexcited frequencies (i.e. $ A(k_{n.exc}) = 0$) in the output spectrum, termed as the \textit{detection lines}, contains valuable information about the 
 system under investigation \cite{RikJohanBook2012} .  

\section{Best Linear Approximation}
\label{Sec:BLA}
\begin{Def}    
\underline{Best Linear Approximation:} The BLA of a nonlinear system is defined as the model $G$ belonging to the set of linear models $\mathcal{G}$, such that \cite{RikJohanBook2012}
\begin{equation}
 G{_{BLA}}(q)= \underset{G(q) \in \mathcal{G}}{\operatorname{arg\,min}} \hspace{0.1cm} \mathbb{E}_u \left( |y(t)-G(q)u(t)|^2 \right)
\end{equation} with $q^{-1}$, the backward shift operator $q^{-1}x(t) = x(t-1)$. The expectation $\mathbb{E}_u(\bullet) $ is taken w.r.t. all signals in the considered signal class.
\end{Def} 
\begin{figure}[!h]
\centering
\includegraphics[width=0.4\textwidth]{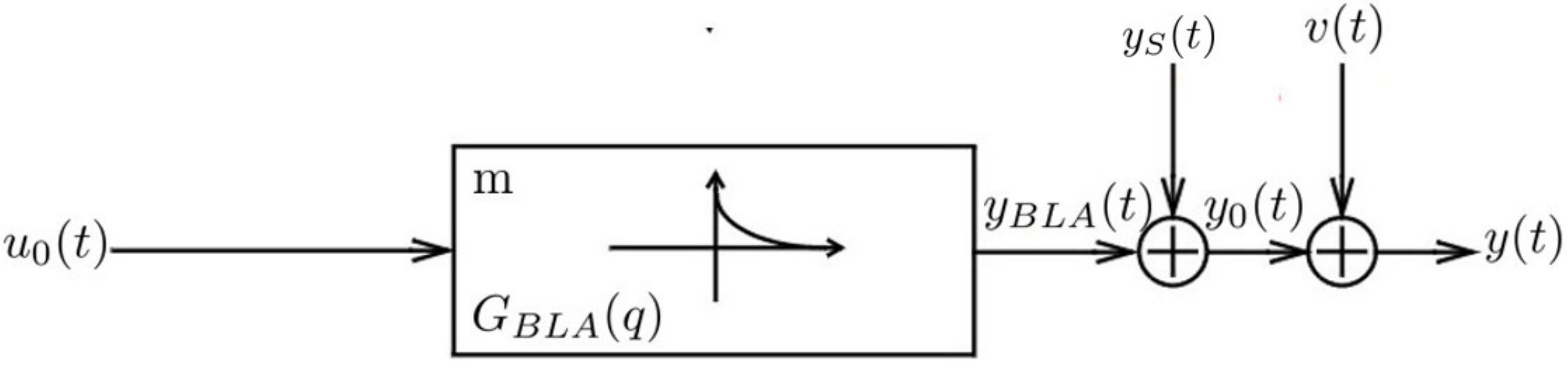}
\caption{ Time domain representation of the problem.}
\label{MultiNonResp}
\end{figure}

\paragraph*{\underline{Set Up}} 
For an infinite length data record $t = -\infty , . .. , N -1,$ the input-output relationship of a discrete-time single-input-single-output (SISO) period-in-same-period out (PISPO) nonlinear system (see Fig. \ref{MultiNonResp}), which is excited with  signals belonging to Definition \ref{def1} can be written as \cite{RikJohanBook2012}:
\begin{equation}
y (t) = G_{BLA}(q)u_0(t) + y_s(t)+ H_0(q)e(t). 
\label{eq:BLA}
\end{equation} 
with $y_s(t)$ the stochastic nonlinear contributions, $u_0(t)$ the exogenous input. The output $y_0(t)$ is disturbed with an additive noise $v(t)$, hence $y (t) = y_0(t) + v(t)$. The noise $v(t)$ is assumed to be filtered white noise, $v(t) = H_0(q)e(t)$, where $H_0(q)$ represents the noise model. For a finite length data record $t = 0, . . . , N -1$,  (\ref{eq:BLA}) must be extended with the initial conditions, or in other words, the transient effects $t_G$, $t_H$ of the dynamic system and the noise filter, respectively:  
\begin{equation}   
y(t) = G_{BLA}(q)u_0(t) + y_s(t)+ H_0(q)e(t)+ t_G(t) + t_H(t). 
\label{TransTermTime} 
\end{equation} Using the definition of discrete Fourier transform (DFT)
\begin{equation}
 X(k)= \frac{1}{\sqrt{N}}\sum\limits_{t = 0}^{N-1}x(t) e^{-j2{\pi}kt/N},
\end{equation} an exact frequency domain formulation \cite{pintelon1997frequency,  RikJohanBook2012} of (\ref{TransTermTime}) is:
\begin{align}
 Y(k) &= G_{BLA}(\omega_{k} ) U_0(k) + Y_s(k) + H_0(\omega_{k}) E(k)\nonumber \\
      &+ T_G (\omega_{k}) + T_H (\omega_{k}) 
 \label{fullLPM}
\end{align} where the index $k$ points to the frequency $kf_s/N$ , with $f_s$  the sampling frequency, and $\omega_{k}=  e^{j2{\pi}kf_s /N} $. The transient terms $t_G(t), t_H(t)$ are described in time domain by rational forms in $q^{-1}$, applied  to  a  delta input, whereas the transient terms $T_G(k) , T_H(k)$ in frequency domain are described by the rational functions in $z^{-1}$, hence they are smooth functions of the frequency. Here, the LPM is utilized to estimate the nonparametric BLA because it makes an optimal use of the smooth behaviour of $G_{BLA}$ and $T_G$ to reduce the leakage errors significantly \cite{ Schoukens2009Nonparam}. As compared to the classical windowing methods it provides a good estimation of the BLA as well as its variance ($\sigma^2_{BLA}$) \cite{Schoukens2009Nonparam}. Other alternatives to estimate the nonparametric BLA are the Fast method, the Robust method \cite{RikJohanBook2012} and the recently developed TRansient Impulse response Modelling Method (TRIMM)\cite{gevers2012transient}.

\section{\textbf{Nonparametric BLA: LPM method}}
\label{NonParam}
In this section, we give a very brief introduction to the LPM method, which is used to estimate nonparametrically the FRF from the input current and the output voltage data. A detailed description and full analysis is also given in \cite{pintelon2010estimation1, pintelon2010estimation2}, 
The basic idea of the LPM method is quite simple:  as stated above the transfer function $G_{BLA}$, and the transient term $T_G$ are smooth functions of the frequency, therefore they can be easily approximated by a complex polynomial in a narrow band of frequency, around a user specified frequency $k$. The parameters of the complex polynomials are directly estimated from the measured input-output data.  Next the estimation of $G_{BLA}(k)$, at any central frequency $k$, is retrieved from this local polynomial model as the measurement of the FRF at that frequency. This procedure is then repeated for all DFT frequencies in the band of interest by shifting the sliding window over one DFT bin. In that way, a local estimate of the FRF is obtained at every frequency.

\subsection{BLA using SISO LPM}
\label{SISOLPM}
From the output error expression described by (\ref{TransTermTime}), and an equivalent relation for the DFT-spectra (\ref{fullLPM}), applied to both the plant $G_{BLA}(q)u_0(t)$ as well as the noise term $v(t)= H_0(q)e(t)$ the output spectrum can be rewritten as:
\begin{align}
 Y(k) &= G_{BLA}(\omega_{k})U_0(k) + T (\omega_{k}) + V_0(k) + Y_s(k)
 \label{eqn:OEE}
\end{align} where $T(\omega_{k}) = T_G(\omega_{k} ) + T_H(\omega_{k} )$, is the generalized transient term that accounts both for the leakage of the plant and noise dynamics.  The remaining noise term is $V_0(k) = H_0(\omega_{k})E(k)$.  It is shown in \cite{RikJohanBook2012} that the contributions $U$, $E$, $Y$ in (\ref{eqn:OEE}) are an $\mathcal{O}(N^0)$, while the transient terms $T_G$ and $T_H$ are an $\mathcal{O}(N^{-1/2})$, where $X = \mathcal{O}(N^p)$ means that for $p < 0$, $\lim_{N \to \infty}\left|\frac{X}{N^p}\right|<\infty$ .

The smoothness of both $G _{BLA}$ and $T$ can be exploited to write the following Taylor series representation, which holds true for the frequency lines $k + r$, with $r = 0, \pm 1, . . . , \pm n.$ 
\begin{align}
 G_{BLA}(\omega_{k+r}) &=  G_{BLA}(\omega_{k})  + \sum\limits_{s = 1}^{R} g_s(k)r^s + \mathcal{O} \left(\frac{r}{N} \right)^{{R+1}} \label{polyLPM1}\\ 
 T(\omega_{k+r}) &= T(\omega_{k}) +  \sum\limits_{s = 1}^{R} t_s(k)r^s + N^{\frac{-1}{2}} \mathcal{O} \left(\frac{r}{N} \right)^{{R+1}}
 \label{polyLPM}
\end{align}
All parameters of $G_{BLA}(\omega_{k})$, $T(\omega_{k})$ and the parameters of the Taylor series $g_s(k),t_s(k), s = 1, . . . , R$, for each frequency line $k$ can be collected into a $2(R + 1)$-column vector $\theta_k$ of unknown complex coefficients defined as 
\begin{equation}
\theta_k \triangleq [G_{BLA}(\omega_{k}) \hspace{0.2em} g_{1}(k) ... g_{R}(k); T{(\omega_k)} \hspace{0.2em} t_{1}(k)... t_{R}(k)]^T,
\label{ThetaParam}
\end{equation}
whereas their respective coefficients are collected in a row vector $K(k, r) $. This allows (\ref{eqn:OEE}) to be rewritten (after neglecting the higher order terms) as:
\begin{equation}
 Y(k + r) = K(R, k+r) \theta_k + V_0(k+r),
 \label{eqn:LPM}
\end{equation}
where $K(R,k + r)$ is a $2(R + 1)$ row-vector, which contains both the structural information, i.e. the powers of $r$ in the polynomial expansions in (\ref{polyLPM1}) and (\ref{polyLPM}) as well as the information about the input signal.  

Now, $2n + 1$ equations (\ref{eqn:LPM}) obtained for $r = 0, \pm 1, . . . , \pm n.$ are then collected into one matrix equation by defining the $(2n + 1)$-vectors $\bar{Y}_{k,n}$ and $\bar{V}_{k,n}$
\begin{align}
\bar{Y}_{k,n} &\triangleq [Y_{k-n} \hspace{0.2em} Y_{k-n+1} . . . Y_k . . .\hspace{0.2em} Y_{k+n-1} \hspace{0.2em} Y_{k+n}]^T \\
\bar{V}_{k,n} &\triangleq [V_{k-n} \hspace{0.2em} V_{k-n+1} . . . V_k . . .\hspace{0.2em} V_{k+n-1}  \hspace{0.2em} V_{k+n}]^T\\
\bar{U}_{k,n} &\triangleq [U_{k-n} \hspace{0.2em} U_{k-n+1} . . . U_k . . .\hspace{0.2em} U_{k+n-1}  \hspace{0.2em} U_{k+n}]^T 
\label{polyLPM2}
\end{align}
This finally results in the following expression
\begin{equation}
 \bar{Y}_{k,n} = K_{k,n}(R,\bar{U}_{k,n}) \theta_k + \bar{V}_{k,n} ,
 \label{LPMFinal}
\end{equation} where the matrix $K_{k,n}(R,\bar{U}_{k,n})$ is a $2(n+1)\times2(R+1)$ matrix. The structure of this matrix is entirely determined by the indices $n$ and $R$ and it contains the input signals $U_{k+r}$ which appear in the input vector $\bar{U}_{k,n}$ defined in (\ref{polyLPM2}). Finally, an estimate of the parameter $\hat{\theta}_k$ is then obtained by solving the following linear least-squares problem:
\begin{equation}
\underset{\theta_k}{\text{min}}[\bar{Y}_{k,n} - K_{k,n}(R,\bar{U}_{k,n})]^H[\bar{Y}_{k,n} - K_{k,n}(R,\bar{U}_{k,n})]
\end{equation} where for any complex vector or matrix $A$, $A^H$ denotes its Hermitian (conjugate) transpose \cite{pintelon2010estimation1}. From (\ref{ThetaParam}), it follows that an estimate of the FRF at the frequency line $k$ is obtained as the first component of the parameter estimate $\hat{\theta}_k:\hat{G}_{BLA}(\omega_k)=\hat{\theta}_k(1)$. 

The condition $n \geq R + 1$ is required between the number of spectral lines in the frequency window around $\omega_k$ and the order of the polynomial approximation, to ensure a full column rank matrix $K_{k,n}(R,\bar{U}_{k,n})$ \cite{RikJohanBook2012}. To reduce the variance of the parameter estimate a larger number of frequencies in the frequency window are taken. In this way, the noise will be averaged over a larger amount of data. Similarly the leakage error decreases with increasing $R$. On the downside, a larger window size results in a larger bias error (or interpolation error).  This is caused by the fact that the transfer function varies over the interval. The smallest interpolation error is obtained for $n = R+1$. A detailed error analysis and the bias-variance trade-off of the LPM is presented in \cite{pintelon2010estimation1, pintelon2010estimation2}.

\section{BLA from Multiple Experiments}
\label{CommBLA}
In this section, we describe two different approaches to use the LPM method described above to estimate the BLA from multiple datasets.

\subsection{Averaging over individual BLAs}
Once the individual estimate of $G_{BLA}$ of a sub-record of data acquired either at a constant or at different operating conditions is available, the estimate of a common BLA ($C_{BLA}$) and its variance can be obtained as explained below. Suppose we carry out $M$ independent experiments either at a sample operating condition or at different settings of SoC, temperature and SoH etc., then the $G_{BLA_{i}}$ of each experiment can be calculated individually using the nonparametric identification procedures described in Section \ref{SISOLPM} above. 

The $C_{BLA}$ of the battery dynamics is calculated from the set of individual BLAs by calculating the sample mean (at each frequency line $k$ in the set of excited frequency lines) of all BLAs. Similarly the variance of $C_{BLA}$ can be obtained by calculating the sample variance of the individual BLAs:
\begin{align}
 C_{BLA}(k) =& \frac{1}{M}\sum_{i=1}^{M}G_{BLA_{i}}(k)\\ 
 {{\sigma}^2}_{C_{BLA}}(k) =& \frac{1}{M-1}\sum_{i=1}^{M}|G_{BLA_{i}}(k)- C_{{BLA}}(k)|^2
 \label{eq:CBLA}
\end{align} 
                               
\subsection{Common BLA using Multi-Input Multi-Output LPM}
\label{MIMOLPM}
Another way of estimating a common BLA of the concatenated data records is by utilising the multi-input multi-output (MIMO) setting of LPM \cite{johan2012conc}. 

\begin{definition}   
\label{Assum1}
\underline{Nonlinear Distortions:} The level of NL are same at different temperatures or levels of SoC w.r.t. the same realisation of the input current load profile.
\end{definition}

\begin{remark}
The NL may be different at different operating conditions (see Section \ref{res}). Nevertheless, we introduce the extended method without loss of generality on the concatenation of two records in the absence of  disturbing  noise and NL;  the  results  apply  to  an  arbitrary  number  of  concatenated  subrecords  in  the  presence  of  disturbing noise and NL.
\end{remark}

Here we consider two data records with lengths $N_1$ and $N_2$ for pedagogical reason but the extension to more datasets is straightforward. For example, for $k=1,2$, we can write $u^{[k]}(t)$ and $y^{[k]}(t)$ with $t = 0,1,...,N{_k}-1$. Consequently, the concatenated data input and output records are then expressed as $u_0 = [u^{[1]}_{0}, u^{[2]}_{0}]$ and $y_0 = [y^{[1]}_{0}, y^{[2]}_{0}]$, respectively. Using (\ref{TransTermTime}) and Assumption \ref{Assum1}, we can write that, 
\begin{align}
 y_0(t) &= [y^{[1]}_{0}, y^{[2]}_{0}]\nonumber \\
  &= G_{BLA}(q)u^{[1]}_{0}(t)+t^{[1]}_G(t) + G_{BLA}(q)u^{[2]}_{0}(t)+ t^{[2]}_G(t)\nonumber \\
  &= G_{BLA}(q)[u^{[1]}_{0}(t), u^{[2]}_{0}(t)] + t_G(t) + t_G(t-N_1)\nonumber \\
  &= G_{BLA}(q)[u_{0}(t)] + t_G(t) + t_G(t-N_1)
 \label{MIMOBLA}
\end{align} For $t<0$, the transient term $t^{[k]}_G(t)= 0$. Similar to (\ref{eqn:OEE}), an equivalent relationship between the input and the output DFTs becomes, where now $\omega_{k} = e^{\frac{2 j\pi k}{N_1+N_2}}$.
\begin{equation}
 Y{_0}(k) = G_{BLA}(\omega_{k} ) U_0(k) + T_G (\omega_{k}) + T_G (\omega_{k})\omega_{k}^{-N_1}
 \label{DFTConc}
\end{equation} It follows from (\ref{MIMOBLA}) and (\ref{DFTConc}), that an additional transient in the concatenation point is added to the output. Another way to write (\ref{MIMOBLA}) is: 
\begin{equation}
 y_0(t) = G_{BLA}(q)u_{0}(t) + G_{T_G}(q)\delta(t) + G_{T_G}(q)\delta(t-N_1)
 \label{DiracEq}
\end{equation} with $\delta(t)$ being a Dirac impulse: $\delta(0) = 1$, and $\delta(t) \neq 0$ if $t \neq 0$.
In (\ref{DiracEq}), the transients are modelled as the response of a linear system to a Dirac impulse in $t = 0$ and in the concatenation point $t = N_1$. The transfer functions $G_{T_G}$ and $G_{BLA}$ have equal denominator. Consequently, we can write (\ref{DiracEq}) as the output of a multiple-input system, that is excited with the concatenated input records at one of the inputs of the system and with Dirac impulses at the beginning of each record that is concatenated ($t = 0$ and $t = N_1$) at the remaining inputs. Hence, the MIMO LPM described in \cite{pintelon2010estimation1} to measure the FRF using concatenated records can be used without any change. 

The major difference in this formulation with  the  SISO  formulation  is  that,  in this particular formulation, the  number  of combined frequencies $2n + 1$ in (\ref{LPMFinal}) will grow with the number of transients. For obtaining an interpolation of order $R$, the number of complex parameters/transient terms which need to be estimated is $R + 1$. Hence, at least $2n + 1 \geq (R + 1)(1 + N_c)$ lines should be combined, with $N_c$ being the number of concatenated subrecords. If the estimation of the  variance of the disturbing noise is also required, then a strict inequality  $2n + 1 > (R + 1)(1 + N_c )$ is needed to have residuals different from zero \cite{johan2012conc}. It follows directly from these inequalities that the interpolation error (bias of the estimate) will increase, while using concatenated sub-records instead of a single data record of the same total length. The reason being that, now the interpolation is made over a larger band of frequency. Similarly, the variance of the estimate will grow because a larger number of parameters are estimated. However, if the record length grows by concatenating data records, the leakage errors are reduced.  

\section{Measurement Setup}
\label{MeasExp} 

In this investigation, the tests are performed on a pre-conditioned battery inside a temperature controlled chamber at different temperatures. A high energy density Li-ion Polymer Battery (EIG-ePLB-C$020$, Li(NiCoMn)) with the following electrical characteristics: nominal voltage $3.65 V$, nominal capacity $20$Ah, AC impedance ($1$ KHz) $< 3m\Omega$ along with the PEC battery tester 
with $24$ channels is used for the data acquisition. 
\subsection{Experiment Design}

An odd-random phase multisine current signal is used to excite the battery within the band of excitation between $1$Hz\textendash$5$Hz. The dynamic range of interest  of the battery for HEVs and EVs applications is covered well within this band of excitation as this frequency bandwidth is corresponding to the bandwidth of the power demand of a vehicle application (acceleration and decelerations), when considering the high power perturbations, as we do here \cite{Firouz2016602}. The selected range also takes into consideration the limitations of the battery tester in terms of the highest sampling frequency. Each period of excitation signal has $5000$ samples and the sampling frequency $f_s$ is set to $50$Hz, which results in a frequency resolution of $f_o = 0.01$Hz. The range of excitation frequency is also limited due to the system limitations of the PEC testers. The input is zero mean with a Root Mean Square (RMS) value of $10$A. A random realisation of the phases of the multisine signal with $7$ periods is acquired at different levels of SoC and temperatures. For the test, using the constant current-constant voltage method, the battery is first charged using a constant $\frac{C}{3}$ rate, where $C$ is the rated capacity, to the maximum charge voltage of $4.1$V. Then, after a relaxation period of $30$ minutes, it is discharged to the desired SoC level Ah-based and considering the actual discharge capacity at  $25$\textdegree C until the end of discharge voltage $3.0$V of the cell. After each discharge, the battery is relaxed for $60$ minutes, before the multisine tests are performed. It is made sure that the synchronisation is maintained between the signal generation and acquisition side. 

\section{Results and Discussion} 
\label{res}
The comparison of the BLA estimation using two methodologies described in Section \ref{CommBLA} is presented here. Two different case studies are discussed: \begin{inparaenum}[(a)] \item  using multiple datasets acquired at the same operating condition \item and using multiple datasets at varying operating conditions of SoC and temperature \end{inparaenum}. 
\subsection{BLA at the same operating condition}
\begin{figure}[!ht] 
 \centering
	\includegraphics[width=0.45\textwidth]{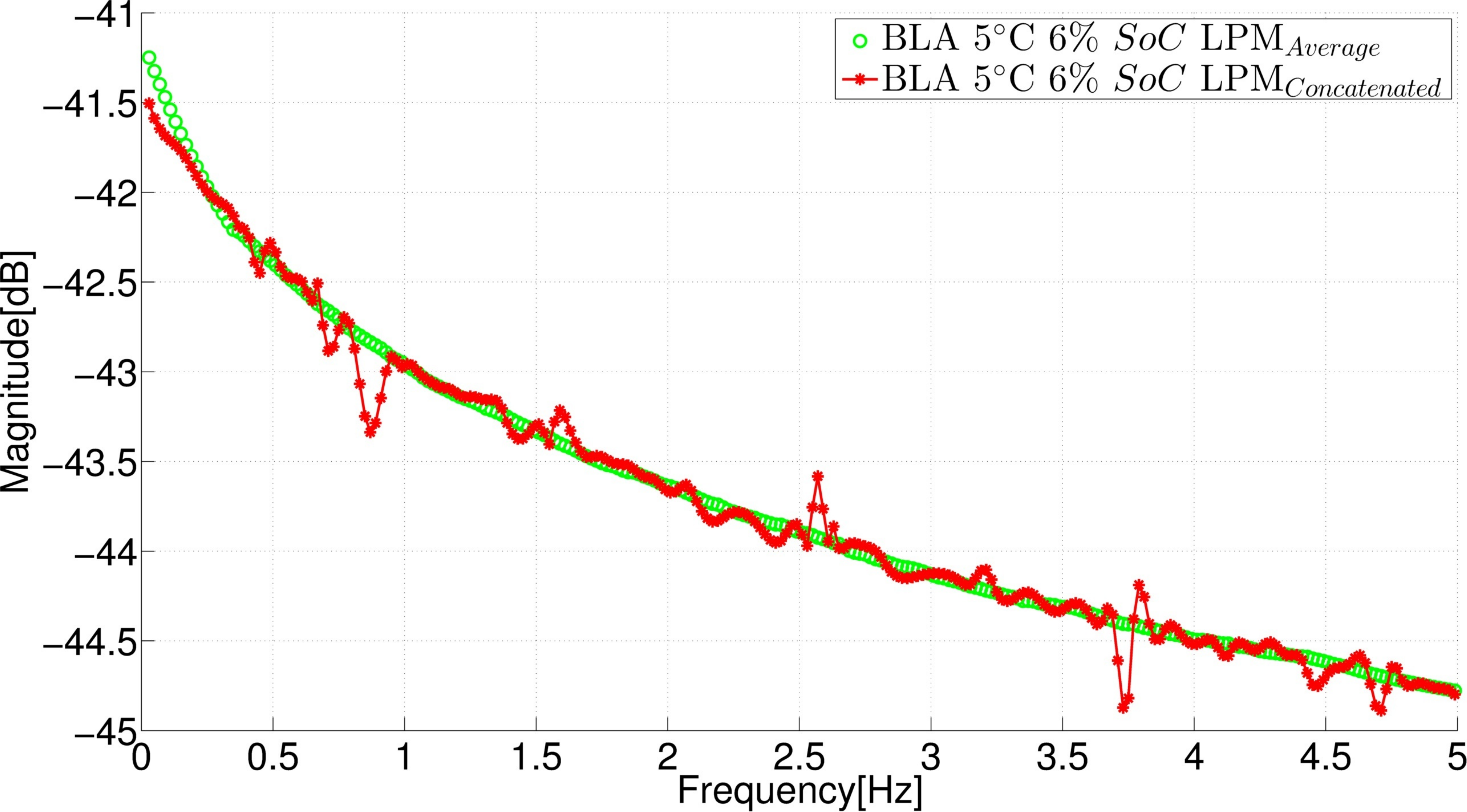}
	\caption{Comparison of the BLAs using LPM averaged (green curve) approach and LPM MIMO approach (red curve) at ($6$\% SoC, $10$A RMS, $5\degree$C).}
	\label{BLA10A5C6SoC}
\end{figure} Figure \ref{BLA10A5C6SoC} shows a comparison between the estimate of BLA using the approaches discussed in Section \ref{CommBLA} from the data from multiple experiments performed at a fixed operating condition of $6$\% SoC, $10$A RMS, $5\degree$C. It can be clearly seen that both approaches result in the estimate of the BLA of the same quality, although the variance of the BLA estimated using the MIMO setting of the LPM is bit larger. 

\subsection{BLA at different operating condition}
Here we present the result of the BLA estimation using the data acquired at different settings of temperatures at a fixed SoC level.
\begin{figure}[ht]
 \centering
	\includegraphics[width=0.45\textwidth]{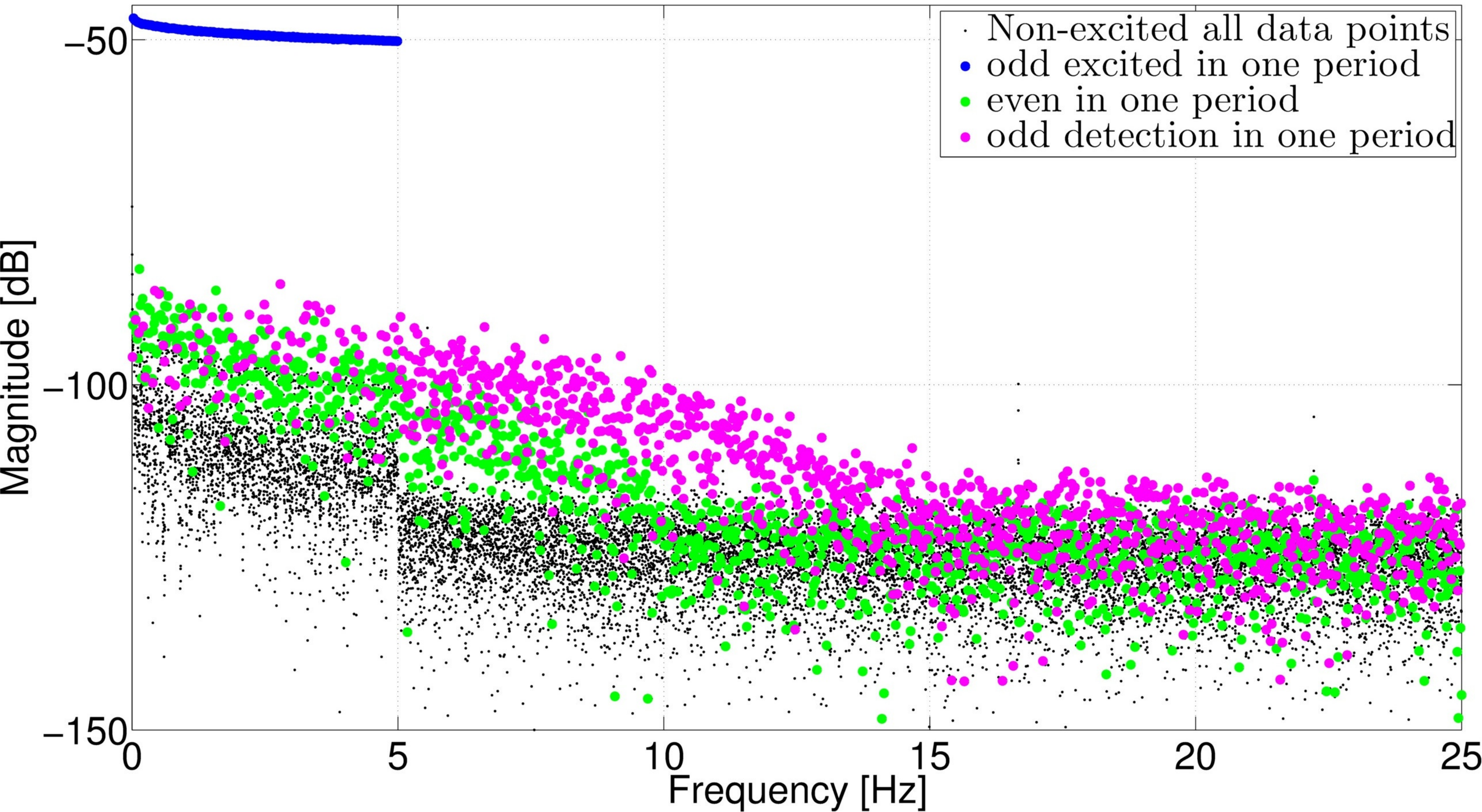}
	\caption{Output voltage response spectrum (blue) at ($10$\% SoC, $10$A RMS, $5\degree$C), Magenta: odd nonlinear distortions, Green: even nonlinear distortions, Black: noise.}
	\label{Resp10A5C10SoC}
\end{figure}
\begin{figure}[ht]
 \centering
	\includegraphics[width=0.45\textwidth]{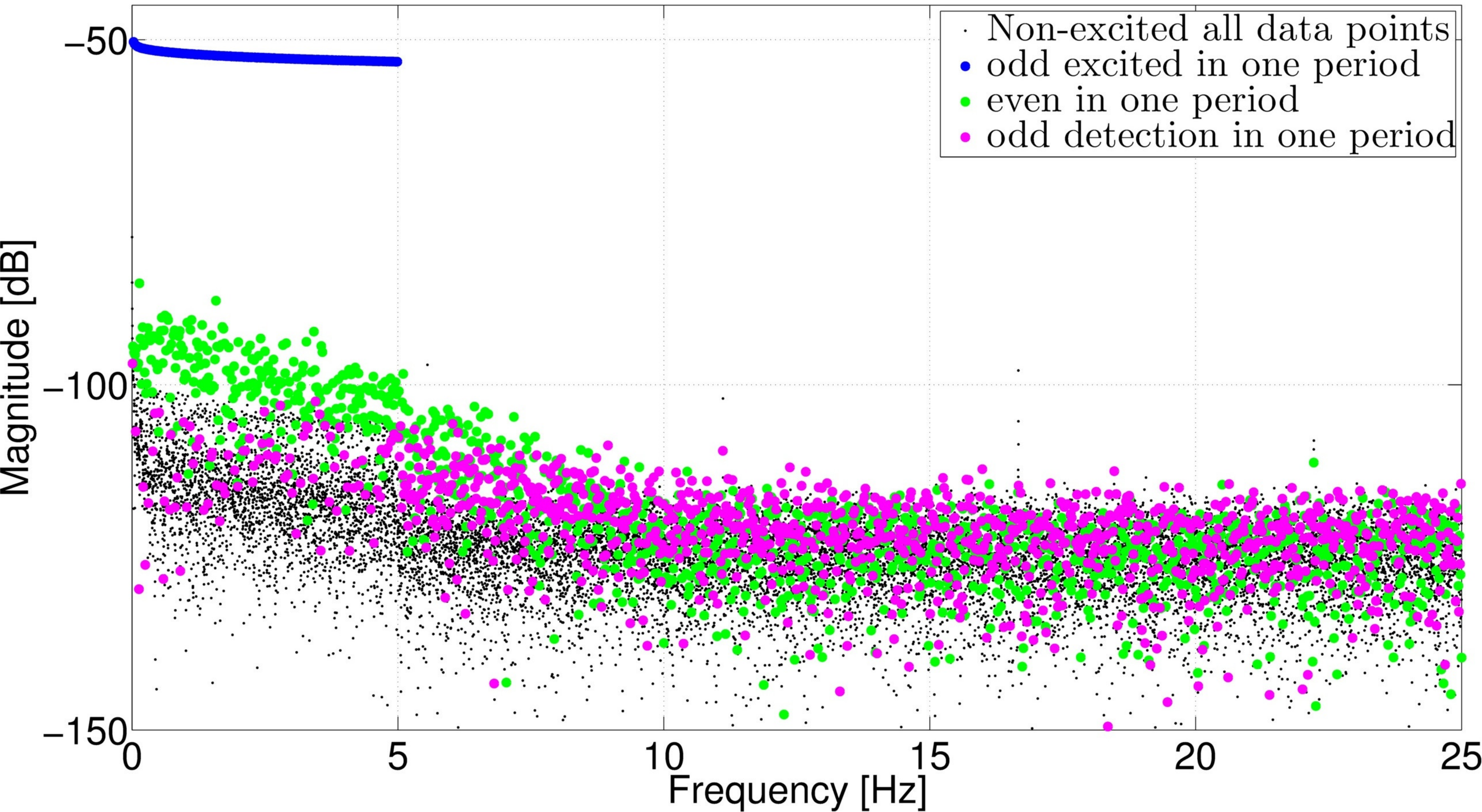}
	\caption{Output voltage response spectrum (blue) at ($10$\% SoC, $10$A RMS, $14\degree$C), Magenta: odd nonlinear distortions, Green: even nonlinear distortions, Black: noise.}
	\label{Resp10A14C10SoC}
\end{figure}
\begin{figure}[ht]   
 \centering
	\includegraphics[width=0.45\textwidth]{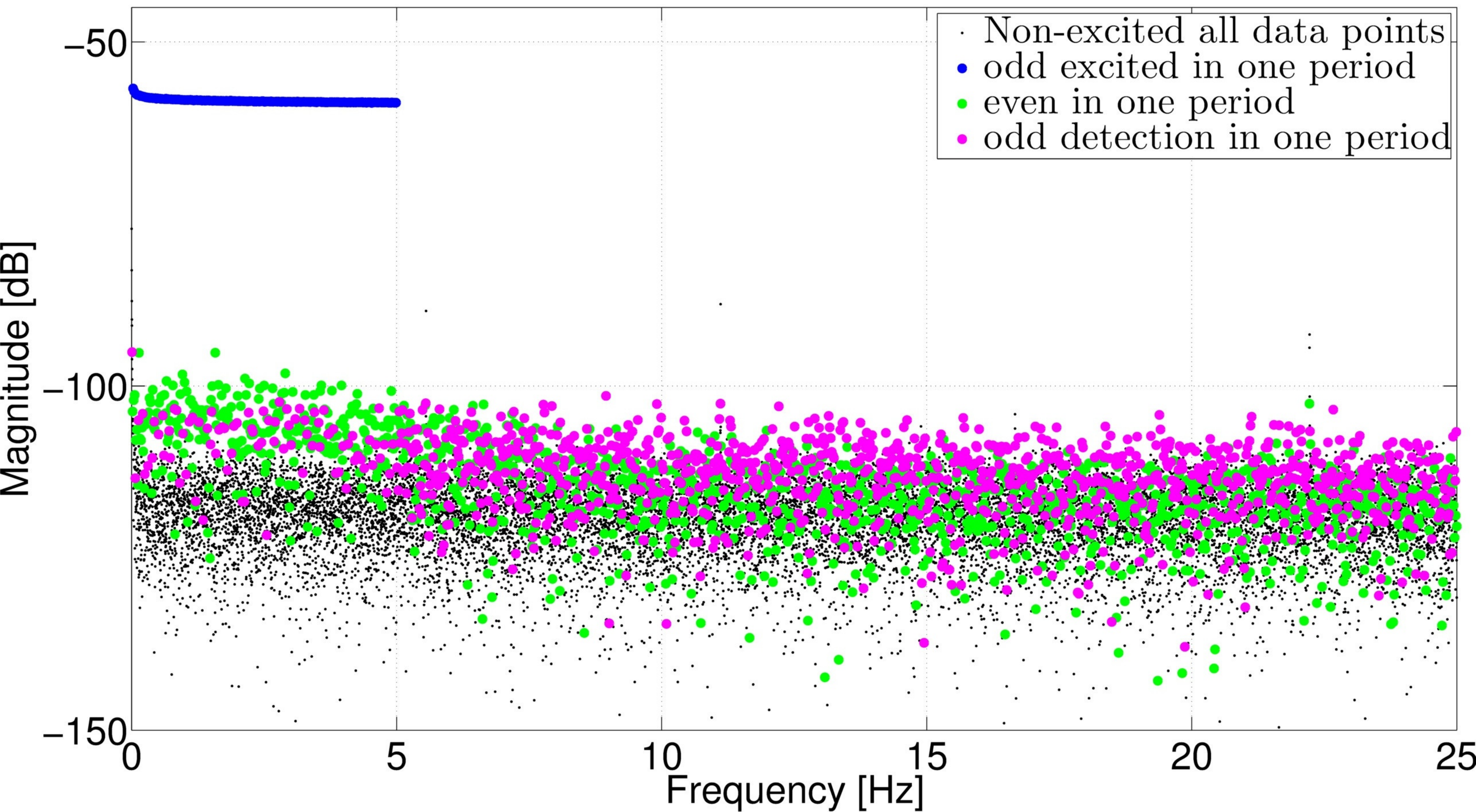}
	\caption{Output voltage response spectrum (blue) at ($10$\% SoC, $10$A RMS, $35\degree$C), Magenta: odd nonlinear distortions, Green: even nonlinear distortions, Black: noise.}
	\label{Resp10A35C10SoC}
\end{figure} \Crefrange{Resp10A5C10SoC}{Resp10A35C10SoC} show the result of nonparametric analysis performed at $10$\% SoC and different settings of temperatures using $10$A RMS multisine input current profile. The readers are referred to \cite{RishiICC2016}, for detailed information on the nonparametric characterization of the battery's short term electrical response. It can be clearly observed from these figures that the level of nonlinear distortions (both even and odd) changes w.r.t. the operating conditions. Hence Assumption \ref{Assum1}, made in Section \ref{CommBLA} is not satisfied. 
\begin{figure}[ht]
 \centering
	\includegraphics[width=0.45\textwidth]{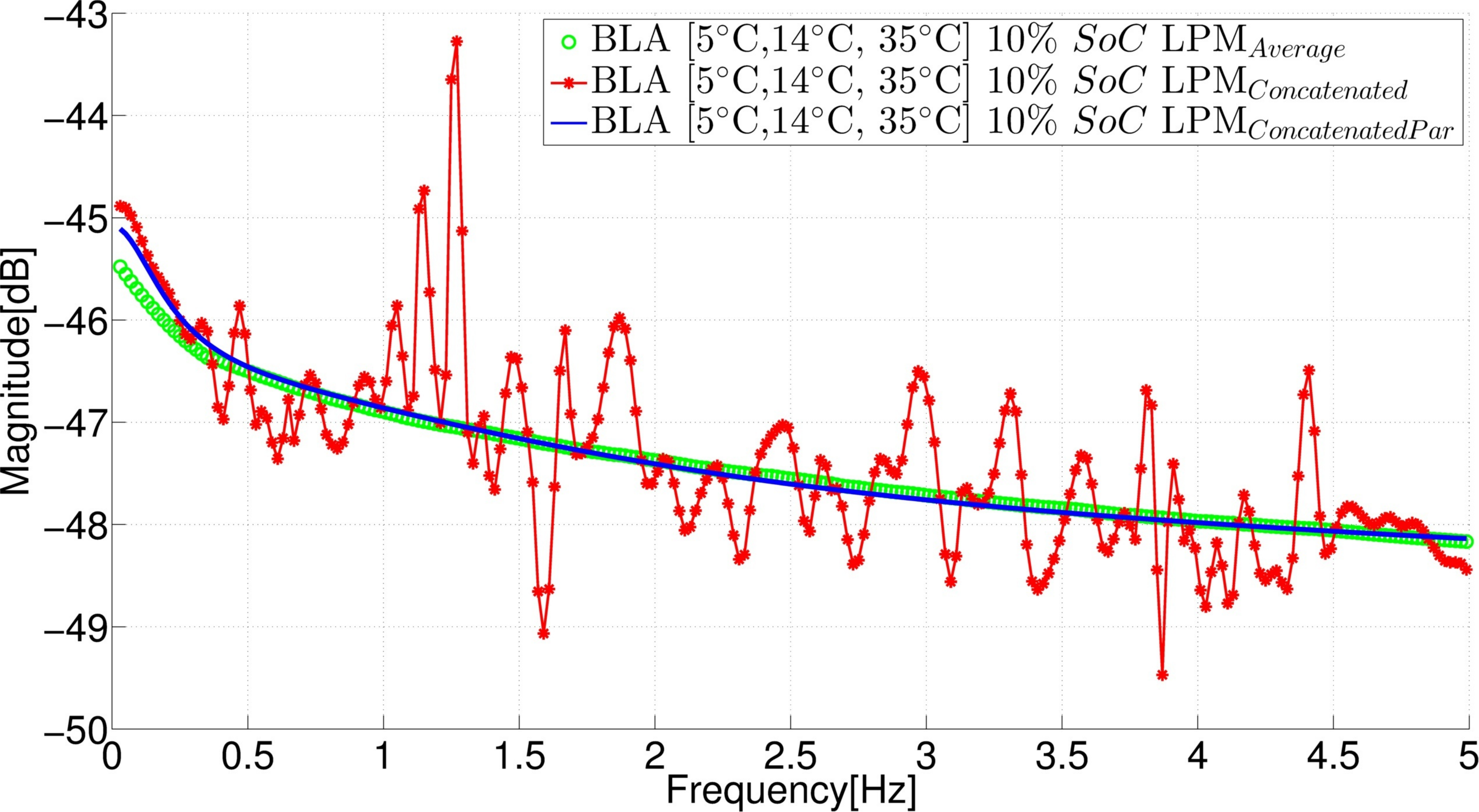} 
	\caption{BLA at ($10$\% SoC, $10$A RMS, [$5\degree$C, $14\degree$C, $35\degree$C]), Comparison between Red: BLA with LPM MIMO settings, Green: LPM averaged, Blue: $3^{rd}-$order parametric model.}
	\label{BLA10A25-40C}
\end{figure} Nevertheless, it can be clearly seen from the Fig. \ref{BLA10A25-40C} that, despite varying levels of nonlinear distortions between different datasets acquired at different temperatures, the BLA estimate using the two approaches is quite similar in magnitude but with a higher variance in the case of LPM MIMO setting. This observation reconfirms the claim made in Section \ref{CommBLA}. 

The final values of the parameters can be used to identify a parametric BLA, which will eventually smoothen the BLA estimate further. A $3^{rd}-$order parametric model is fitted on the nonparametric BLA by solving the nonlinear weighted least squares cost function \cite{RikJohanBook2012} in frequency domain (see blue curve in the Fig. \ref{BLA10A25-40C}, a $3^{rd}-$order discrete-time transfer function fitted on the BLA estimated using MIMO LPM setting at different temperatures). The frequency domain approach allows us to put user-defined weighting in the frequency band of interest by exploiting the calculated noise variance during the estimation of nonparametric BLA. A range of model orders were evaluated and the order of the final parametric model was determined using a signal theoretic measure such as the minimum description length (MDL) criterion (see page no. $439$ of \cite{RikJohanBook2012}). Hence, individual BLAs estimated at varying operating points can be used to develop black-box linear time-varying or parameter-varying models or the C$_{BLA}$ can be used as initialization for the nonlinear model structure proposed in \cite{RishiTCST2016}.

\section{Conclusion}
\label{conc}
This paper proposed LPM based approaches to estimate nonparametrically the BLA of the battery's short term electrical dynamics from multiple datasets. The proposed framework paves the way for handling data records of arbitrary lengths acquired under similar or different conditions and dealing with nonlinear distortions efficiently. This gives a practical advantage when performing longer experiments is either not feasible or rather expensive and time consuming. Similarly, the data of extremely poor quality can also be handled. This whole process can be carried out in relatively short measurement time due to the use of broadband excitation signals for identification.

\section*{ACKNOWLEDGMENT}
\footnotesize {This work was supported in part by the IWT-SBO BATTLE 639,  Fund for Scientific Research (FWO-Vlaanderen), by the Flemish Government (Methusalem), the Belgian Government through the Inter university Poles of Attraction (IAP VII) Program, and by the ERC advanced grant SNLSID, under contract 320378.}
\bibliographystyle{IEEEtran}
\bibliography{IEEEtranBibNew} 

\end{document}